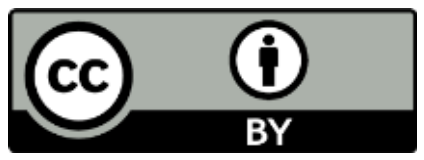

# Hybrid Approach to Directed Fuzzing

[2] *D.A. Parygina, ORCID: 0000-0002-4029-0853 <parygina.darya.2001@yandex.ru>*
[1,2] *T.P. Mezhuev, ORCID: 0009-0009-0610-287X <mezhuevtp@ispras.ru>*
[1] *D.O. Kutz, ORCID: 0000-0002-0060-8062 <kutz@ispras.ru>*

[1] *Ivannikov Institute for System Programming of the Russian Academy of Sciences,*
*25, Alexander Solzhenitsyn st., Moscow, 109004, Russia.*

[2] *Lomonosov Moscow State University,*
*GSP-1, Leninskie Gory, Moscow, 119991, Russia.*

**Abstract.** Program analysis and automated testing have recently become an essential part of SSDLC. Directed greybox fuzzing is one of the most popular automated testing methods that focuses on error detection in predefined code regions. However, it still lacks ability to overcome difficult program constraints. This problem can be well addressed by symbolic execution, but at the cost of lower performance. Thus, combining directed fuzzing and symbolic execution techniques can lead to more efficient error detection.
In this paper, we propose a hybrid approach to directed fuzzing with novel seed scheduling algorithm, based on target-related interestingness and coverage. The approach also performs minimization and sorting of objective seeds according to a target-related information. We implement our approach in Sydr-Fuzz tool using LibAFL-DiFuzz as directed fuzzer and Sydr as dynamic symbolic executor. We evaluate our approach with Time to Exposure metric and compare it with pure LibAFL-DiFuzz, AFLGo, and other directed fuzzers. According to the results, Sydr-Fuzz hybrid approach to directed fuzzing shows high performance and helps to improve directed fuzzing efficiency.

**For citation:** Parygina D.A., Mezhuev T.P., Kutz D.O. Hybrid Approach to Directed Fuzzing. Trudy ISP RAN/Proc. ISP RAS, vol. 37, issue 6, part 2, 2025, pp. 53-64. DOI: 10.15514/ISPRAS-2025-37(6)-19.

**Acknowledgements**. The results were obtained using the services of the Ivannikov Institute for System Programming (ISP RAS) Data Center.

# Гибридный подход к направленному фаззингу

[2] *Д.А. Парыгина, ORCID: 0000-0002-4029-0853 <parygina.darya.2001@yandex.ru>*
[1,2] *Т.П. Межуев, ORCID: 0009-0009-0610-287X <mezhuevtp@ispras.ru>*
[1] *Д.О. Куц, ORCID: 0000-0002-0060-8062 <kutz@ispras.ru>*

[1] *Институт системного программирования им. В.П. Иванникова РАН,*
*Россия, 109004, г. Москва, ул. А. Солженицына, д. 25.*

[2] *Московский государственный университет имени М.В. Ломоносова,*
*Россия, 119991, Москва, Ленинские горы, д. 1.*

**Аннотация.** Анализ программ и автоматизированное тестирование в последнее время стали неотъемлемой частью РБПО. Направленный фаззинг – один из самых популярных методов автоматизированного тестирования, который фокусируется на поиске ошибок в заранее определенных областях кода. Однако этот метод не способен преодолевать сложные программные ограничения. Эта проблема может быть эффективно решена с помощью символьного выполнения, но ценой более низкой производительности. Таким образом, комбинирование методов направленного фаззинга и символьного выполнения может привести к более эффективному поиску ошибок в программах.

В этой статье мы предлагаем гибридный подход к направленному фаззингу с оригинальным алгоритмом планирования входных данных, основанным на пользе для достижения целевых точек и увеличения покрытия кода. В подходе также выполняется минимизация и сортировка результатов анализа в соответствии с информацией о целевых точках. Мы реализовали наш подход в инструменте Sydr-Fuzz, используя LibAFL-DiFuzz в качестве направленного фаззера и Sydr в качестве динамического символьного исполнителя. Мы оценили наш подход с помощью метрики Time to Exposure и сравнили его с чистым LibAFL-DiFuzz, а также с инструментом AFLGo и другими направленными фаззерами. Согласно результатам, гибридный подход Sydr-Fuzz к направленному фаззингу демонстрирует высокую производительность и помогает повысить эффективность направленного фаззинга.

**Ключевые слова:** направленный фаззинг; символьное выполнение; гибридный фаззинг; анализ программ.

**Для цитирования:** Парыгина Д.А., Межуев Т.П., Куц Д.О. Гибридный подход к направленному фаззингу. Труды ИСП РАН, том 37, вып. 6, часть 2, 2025 г., стр. 53–64 (на английском языке). DOI: 10.15514/ISPRAS–2025–37(6)–19.

**Благодарности:** Результаты получены с использованием услуг Центра коллективного пользования Института системного программирования им. В.П. Иванникова РАН – ЦКП ИСП РАН.

## *1. Introduction*

Program analysis and automated testing have recently become an essential part of SSDLC [1-3]. New software should be properly tested before releasing to reduce the probability of vulnerabilities.

Among various testing methods, fuzzing [4-5] still remains one of the most popular choices. It is a comparatively fast dynamic testing technique that allows checking program behavior on a wide range of input seeds. However, for certain specific goals, such as static analysis report verification, patch testing, crash reproduction, coverage-guided greybox fuzzing may appear less effective as its only metric to optimize is the global coverage value. Directed greybox fuzzing [6], in opposite, focuses on predefined program locations, called target points, to address these goals. In directed fuzzing, there are special proximity metrics that show how close a certain program execution is to reaching target points.

But for all the advantages of directed fuzzing approach, it still lacks the ability to overcome difficult program constraints, and thus can be unable to explore important code parts. Such problems are well addressed by symbolic execution [7-9] technique that interprets program instructions in terms of symbolic values associated with program variables, and discovers new program paths by inverting branch constraints. SMT-solvers [10-11] assistance allows symbolic executors solving complicated formulas and generate inputs that will lead the program along the desired execution path.

The main limitation of symbolic execution is its high overheads. Thus, both directed greybox fuzzing and symbolic execution can benefit from their combination. In this paper we propose hybrid directed fuzzing approach with novel seed exchange algorithm. We propose also post-analysis minimization and sorting of objectives (generated inputs considered to be analysis results) to determine which target points are reached, and what side effects (e.g., real program crashes or timeouts) are obtained.

This paper makes the following contributions:

- We propose a hybrid approach to directed fuzzing with novel seed scheduling algorithm, based on target-related interestingness and coverage. The approach also performs minimization and sorting of objective seeds according to a target-related information.
- We implement our approach in Sydr-Fuzz [12] tool using LibAFL-DiFuzz [13] as directed fuzzer and Sydr [9] as dynamic symbolic executor.
- We evaluate our approach with Time to Exposure metric and compare it with pure LibAFL-DiFuzz, AFLGo [6], BEACON [14], WAFLGo [15], WindRanger [16], FishFuzz [17], and Prospector [18]. We get 3 out of 7 best results with speedup up to 1.86 times from the second-best results. We get significant improvements on 3 out of 7 examples comparing with pure LibAFL-DiFuzz fuzzer with gaining speedup up to 4.9 times, and reaching an undiscovered target point.

## *2. Related Work*

### 2.1 AFLGo

One of the first solutions in the field of directed greybox fuzzing was AFLGo [6] tool. The authors solve the problem of exploring given target points in code by scheduling energy of the inputs based on *simulated annealing* algorithm [19-20]. According to the algorithm, during energy scheduling, large energy values are assigned to inputs that provide the closest approximation to the target points.

AFLGo treats the target point exploration problem as an optimization problem. The resulting set of inputs with the largest energy values converges asymptotically to a set of global optimal solutions that allow to explore the program execution paths closest to the target points. To regulate the convergence rate, the authors use the exponential law of decreasing parameter *temperature*, which characterizes the probability of assigning large energy values to the worst solutions.

A metric based on the distances over the target basic blocks is proposed as a metric for the proximity of the program execution to reach the target points. Before fuzzing begins, a stage of static program instrumentation is performed: a program call graph and intra-procedural control flow graphs (CFGs) are constructed. Each vertex is matched with a metric value characterizing the distance from this vertex to all target vertices of the graph. The metric value for specific input is calculated as a normalized sum of metric values of each basic block in the program execution trace. The proposed metric allows to assign large energy values to those inputs that provide a shorter distance to the target points. However, such a solution has some drawbacks. First of all, due to the need for calculation metric values for all CFG basic blocks and all program functions, the stage of static instrumentation entails considerable time expenses and a large number of unnecessary calculations. At that, any changes in the target program require the static instrumentation stage to be performed anew. In addition, the solution converges to the selection of input data providing the minimum path length to the target positions without considering the execution context, indirect transitions, and program data flow.

### 2.2 BEACON

BEACON [14] uses *reachability analysis* to prune irrelevant basic blocks. It computes the reachability to the target point for every basic block and then the irrelevant ones are pruned, so they are not being instrumented. Interesting blocks are used in further analysis to apply constraints on them. Such approach tends to reduce overheads significantly, however limits BEACON to support only single-target mode.

In the next phase the backward interval analysis is used. It slices the program using statically computed control flow information and then applies the analysis that builds value intervals for all variables in program basic blocks. These value intervals are used to construct formulas for target reachability depending on value constraints. Then it inserts such formulas in code as instrumentation. At runtime values are checked for these constraints, if so, then program proceeds further to target point, otherwise it terminates.

## 2.3 WAFLGo

WAFLGO [15] is a directed greybox fuzzer that was designed to discover bugs presented in new code commits. Traditional fuzzers usually focus only on reaching the commit-modified code, but they miss bugs that emerge in the wider affected code. WAFLGO solves this by introducing *Critical Code Guidance*, which separates *path-prefix code* (that leads to the change site) from *data-suffix code*, which represents the code influenced by the modified variables. It also has multi-target scenarios to balance precision and efficiency using a lightweight, function-level distance metric, effective input generation strategies using special mutation stages. By focusing the fuzzing effort on the critical code introduced in a commit, WAFLGo improves the efficiency and effectiveness of bug finding. It allows developers to quickly identify and fix vulnerabilities introduced by recent changes, reducing the risk of shipping buggy code.

## 2.4 WindRanger

WindRanger [16] is a directed greybox fuzzer that was designed to improve the efficiency of vulnerability discovery toward specific code regions (called *target sites*). Many greybox fuzzers treat all basic blocks the same, but WindRanger introduces the concept of *Deviation Basic Blocks* (*DBB*), the blocks that have the strongest impact on fuzzing planning, using them to calculate seed distances, guide mutations, and prioritize inputs more accurately. It also uses data flow analysis (*taint tracking*) to evaluate the difficulty of reaching conditions, enhancing both precision and efficiency. By actively identifying and mitigating path divergence, WindRanger improves the efficiency and effectiveness of DGF. It reduces the amount of time wasted exploring irrelevant code regions, allowing the fuzzer to focus on paths that are more likely to lead to the target and uncover vulnerabilities.

## 2.5 FishFuzz

FishFuzz [17] addresses the challenge of generating high-quality inputs for directed greybox fuzzing, specifically focusing on the creation and adaptation of a dictionary of "interesting" input tokens. It introduces an *adaptive dictionary creation* and update strategy. Instead of relying solely on seed inputs, it monitors the program's execution during fuzzing and extracts constant values used in comparisons (e.g., equality checks, greater-than/less-than). These values are then dynamically added to the dictionary. This allows FishFuzz to "fish" for relevant constants revealed during the fuzzing process itself. FishFuzz can optionally integrate with a constraint solver (Z3) to further refine and generalize the extracted constants, potentially discovering related values that can lead to new code coverage.

## 2.6 Prospector

Prospector [18] tackles the problem of imprecise call site targeting in directed greybox fuzzing. It introduces a novel approach that combines *precise call site identification* with *adaptive seed selection* for directed greybox fuzzing. Prospector uses a static analysis technique that is more sensitive to the program's control flow and data flow context. This allows it to differentiate between different call sites of the target function by analyzing the code paths that lead to them. To further refine the call site identification, Prospector performs dynamic validation during fuzzing. It monitors the execution of the program and confirms whether the execution path actually leads to the intended call site based on input. Prospector uses the information from the static and dynamic analysis to prioritize seed inputs that are more likely to reach the correct call site. It ranks seed inputs based on their estimated distance to the target call site (considering the context information).

## 2.7 Hybrid directed fuzzing

In order to help directed fuzzer overcome complicated code constraints, several works propose a hybrid approach. The fuzzer is aided by symbolic or concolic executor that generates new inputs and discovers new code coverage.

In 1dVul [21], a dominator-based symbolic execution mechanism is involved when the fuzzer gets stuck for a specified time threshold. It takes the input with the shortest distance to the target point and tries to generate new ones that could reach target point dominators on the further paths.

DrillerGO [22] limits directive concolic execution to a subset of specific branch constraints. This subset is defined during backward pathfinding analysis that is based on searching suspicious API call strings in the recovered control flow graph.

Berry [23] and LeoFuzz [24] run concolic execution in parallel with the directed fuzzer sharing inputs via several input queues. The division is made based on inputs priority and capability of reaching target points, or discovering new coverage.

HyperGo [25] proposes an Optimized Symbolic Execution Complementary scheme for hybrid directed fuzzing. The scheme prunes the unreachable unsolvable branches, and prioritizes symbolic execution of the seeds whose paths are closer to the target and have more branches that are difficult to cover with the fuzzer.

Despite the variety of approaches, all of them are closed-source with no way to conduct comparison experiments.

## *3. Hybrid approach to directed fuzzing*

In this paper, we present a novel hybrid approach to directed fuzzing. We combine directed greybox fuzzing and symbolic execution to share interesting inputs and benefit from each other. In this section, we describe the details of our integration approach.

## 3.1 Architecture overview

The overall integration scheme is shown in Fig. 1.

There are instances of directed greybox fuzzer and symbolic executor working in parallel and controlled by orchestrator instance. Orchestrator manages the whole analysis pipeline (described in section 4.1), validates configuration, launches fuzzer and symbolic executor instances, and provides their effective communication. After analysis, orchestrator sorts and minimizes found objectives – inputs that allow reaching specified target points.

Directed fuzzing tool LibAFL-DiFuzz [13] performs greybox fuzzing of the program focusing on reaching predefined target points. LibAFL-DiFuzz calculates proximity metric for each target program execution. Proximity metric is based on similarity between Enhanced Target Sequences (ETS) and the execution trace. ETS are built separately for each target point, and contain all the dominator functions and basic blocks of this target point.

Symbolic executor tool Sydr [9] performs concolic execution of the program. During the concrete execution of the tested program on some input, Sydr collects symbolic path constraints. They are used to invert branches by building corresponding path predicates and validating them with SMT-solver [11]. For all successful solutions new inputs are generated.

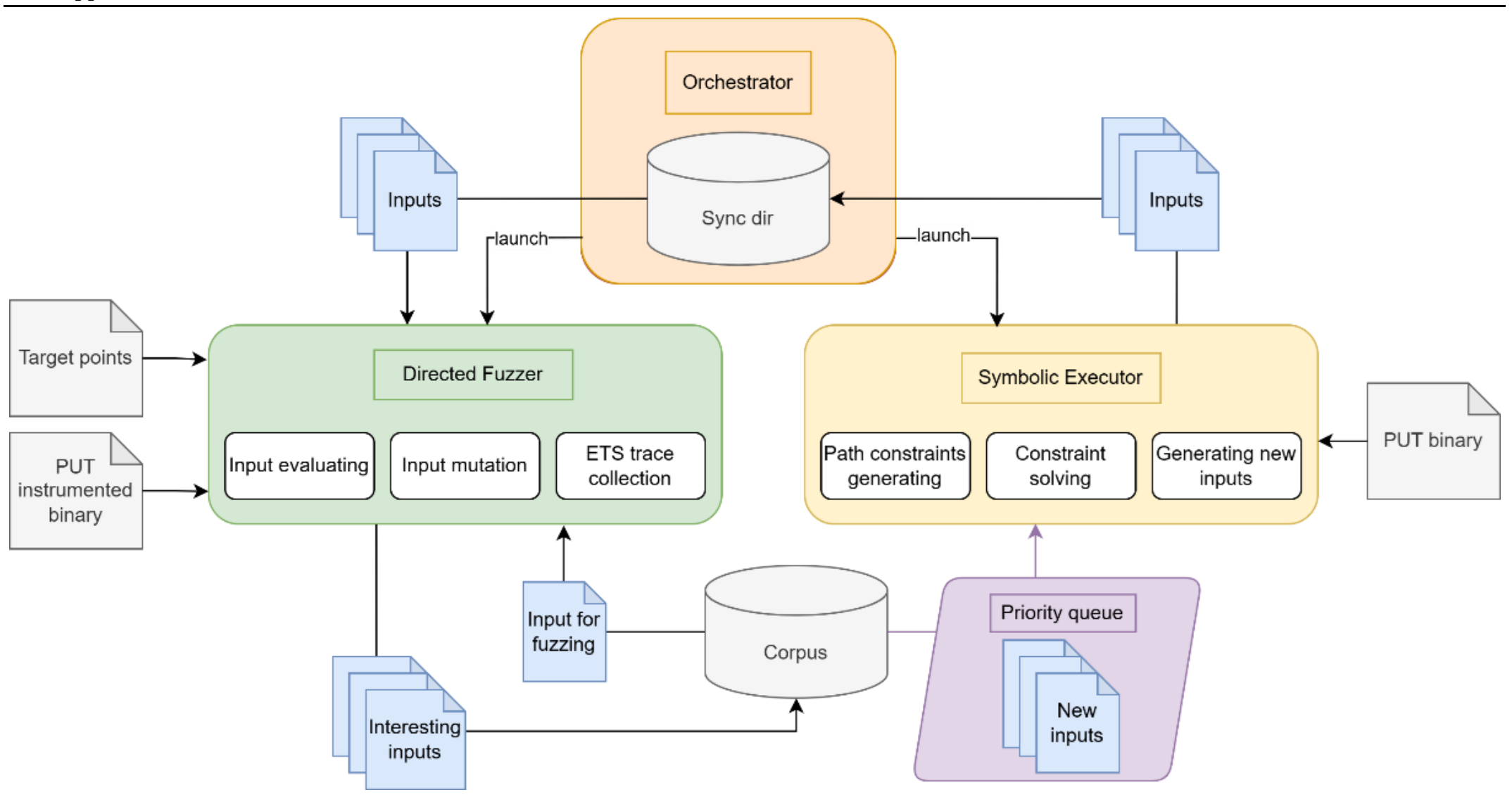

*Fig. 1. Hybrid directed fuzzing scheme.*

## 3.2 Seed scheduling algorithm

In hybrid fuzzing, fuzzer and symbolic executor are running in parallel and sharing input seeds. Symbolic executor puts generated inputs in a separate directory called *sync dir*. LibAFL-DiFuzz performs synchronization with symbolic executor at dynamically calculated time intervals by analyzing all seeds from this directory. It measures whether each seed is interesting for ETS metric or explores new coverage, and if either is true, saves it into the fuzzing corpus directory *corpus*. In case an input causes the program to crash or hang, or leads it to one of the specified target points, it is saved into an *objective directory*. Additionally, LibAFL-DiFuzz has a feature to import all seeds from symbolic executor without preliminarily measuring them. In some cases, the fuzzer cannot produce many inputs on specific target programs, thus adding all seeds can give a boost to find new coverage paths or improve ETS metric.

Scheduling inputs for symbolic execution is another part of seed exchanging algorithm. Since Sydr analyzes only one program execution path at a time, the quality of selected seed is crucial for hybrid fuzzing performance. All seeds generated by the fuzzer and stored in its *corpus* directory are ordered by the orchestrator with the help of *priority queue* (see section 4.2). The seeds are ranked by *interest* score, which is calculated based on ETS exploration and coverage growth. The input is assigned higher priority if it is useful for both ETS and coverage exploration. After that, the seed is taken from the queue and analyzed by symbolic executor. In case of successful path inversion, the modified input is put into the *sync dir*.

Synchronization interval for the fuzzer is determined dynamically to improve fuzzing performance. Default time is 60 seconds, but after the first synchronization it is computed as *max(60,* ***t*****3)*, where ***t*** is the time in seconds that was spent on importing all seeds from symbolic executor. This heuristic is needed to compensate a possibly long synchronization and allocate more processor time on fuzzing process.

## 3.3 Objective sorting and minimization

In directed fuzzing, an input that is considered objective can be one of three types: 1) input that reaches one or more target points without crashing the program; 2) a real crash; 3) a timeout (set to 10 seconds by default). For each objective its type is explicitly specified in the corresponding metadata file created by LibAFL-DiFuzz (see section 4.3). The list of reached target points is also saved in this metadata file.

Once the fuzzing is complete, all found objectives can be minimized and sorted by analyzing their metadata files. Objective minimization is based on ETS novelty. During the target program execution on certain input data, the fuzzer collects an ETS trace – a sequence of IDs of visited basic blocks belonging to at least one ETS. This trace is also saved to the corresponding metadata file. If several objectives have the same ETS traces, we can keep only one of them since only unique ETS traces can bring new useful feedback. Thus, all objectives can be clustered by ETS traces, and only one from each cluster can be taken. This will result in a minimized objective set with variety of ETS traces.

After minimization part is done, remaining objectives are sorted by reached target points. We create a separate directory for every reached target point and copy all corresponding objectives there.

## *4. Implementation*

We implemented our hybrid approach to directed fuzzing with Sydr-Fuzz [12] tool as orchestrator, LibAFL-DiFuzz [13] as directed fuzzer, and Sydr [9] as dynamic symbolic executor. In this section, we describe practical details of the integration.

### 4.1 Sydr-Fuzz pipeline

Sydr-Fuzz tool allows to run hybrid fuzzing by launching such fuzzers as libFuzzer [1], AFL++ [2], or Honggfuzz [26] alongside with Sydr symbolic executor. For directed fuzzing we implemented a hybrid fuzzing approach that allows Sydr-Fuzz combining LibAFL-DiFuzz and Sydr.

Analysis specification is set by configuration TOML-files that match the following sample:

```
[sydr]
args = "-j 4"
target = "/target_sydr @@"
jobs = 3

[difuzz]
target = "/target_libafl_difuzz @@"
args = "-j2 -i /corpus -e /ets.toml"
path = "/fuzzer_libafl_difuzz"
```

This sample Sydr-Fuzz configuration specifies two tables – **[sydr]** and **[difuzz]**, containing parameters for launching each tool, respectively. Every table specifies arguments to run a tool, and a binary program to analyze. Symbolic executor Sydr requires the program to be built without any instrumentation (except debugging information *-g*). For directed fuzzer LibAFL-DiFuzz the program should be compiled with specific fuzzer instrumentation that allows tracking ETS traces. Directed fuzzer also needs its own configuration *config.toml* that specifies target points. To improve analysis performance, Sydr-Fuzz can also launch several instances of each tool in parallel (*jobs* parameter).

Once Sydr-Fuzz has been started with the specified configuration, it launches the fuzzer and verifies that all parallel clients specified by *jobs* parameter have been successfully started. It also enables *watchdog* to monitor a fuzzing *corpus* and track new files. After that, the actual fuzzing stage begins. There are four main tasks performed by Sydr-Fuzz during the hybrid directed fuzzing with LibAFL-DiFuzz and Sydr:

1. Sydr-Fuzz constantly keeps launching the specified number of Sydr workers. For each worker the most interesting seed is pulled from the *priority queue* (see section 4.2). All inputs generated by Sydr are stored into *sync dir* that is analyzed by LibAFL-DiFuzz directly.

2. Every second Sydr-Fuzz parses a fuzzer log and prints a basic information from every client: the number of files in its corpus, the number of objectives found, execution speed, etc.
3. Every minute Sydr-Fuzz updates the *priority queue* with new files from fuzzing *corpus,* which are received from *watchdog*. It also checks conditions to stop analysis, such as time without new coverage, or reaching all target points. Sydr-Fuzz prints analysis statistics (total number of files in the corpus, total number of objectives, average execution speed, last saved objective/corpus time, etc.), and the list of unique reached target points.
4. When Sydr-Fuzz receives a signal to terminate, all LibAFL-DiFuzz clients and Sydr-workers are killed, the final statistics are printed. After that, all files from the *objective directory* are minimized and sorted according to the algorithm described in section 3.3.

## 4.2 Seed priority queue

The *priority queue* is organized as a binary heap, where the seeds are stored along with a special data structure describing input priority. This structure has three fields:

1) *is_interestring_ets* flag shows whether the input is interesting for ETS;
2) *is_interesting_map* flag shows whether it is interesting for program coverage;
3) *file score* metric based on file system metadata: file creation time / file size. Let us call the first two flags *ETS score* and *coverage score*, respectively.

At runtime, Sydr-Fuzz collects a list of newly found inputs from the *corpus* once a minute. For each of them, *ETS* and *coverage scores* are obtained from LibAFL-DiFuzz metadata described in section 4.3. The *file score* is calculated based on file system metadata, and a final structure with three fields is constructed. Since the *priority queue* is a binary heap, when a new seed is added, it is placed in the correct sorted order. That is, when scheduling inputs for symbolic execution, the most priority is given to seeds that discover new ETS traces (interesting for directed fuzzing). The next priority is assigned to seeds that discover new program coverage. A *file score* is a metric that allows choosing new file according to its size. While it is important to always choose the newest file in order to produce the most relevant inputs for the fuzzer, symbolic executor may be less effective on large inputs.

## 4.3 LibAFL-DiFuzz seed metadata

For the effective communication between LibAFL-DiFuzz and Sydr, all important input information is saved in a separate metadata file next to the input. This file includes fields used for calculating seed priority when adding new seed to Sydr *priority queue*, and also fields used for crash sorting and minimization.

Seed priority is based on *ETS score* and *coverage score* (see section 4.2) that are saved by LibAFL-DiFuzz in the form of flags. Whether the seed is interesting for ETS or program coverage, is determined when evaluating LibAFL modules ETSFeedback and MapFeedback after target program execution, respectively.

Objective minimization is based on ETS trace values that are also saved in metadata files. ETS trace is kept as a list containing IDs of visited basic blocks belonging to at least one ETS. And for sorting, we need to know which target points are reached during execution on each objective file. The list of reached target points is saved to metadata file too, and each list element keeps source file name and line number of the target point. For the further analysis and results presentation, we need to know whether the objective is a real crash, or timeout, or just the one that leads to reaching some target points. For these reasons, two flags are saved in metadata: *is_crash* and *is_timeout*. They are set when evaluating LibAFL modules CrashFeedback and TimeoutFeedback, respectively. All seeds in *objective directory* with both flags set to false are considered to be just reaching target points.

## 5. Evaluation

We evaluated our hybrid approach using TTE (Time to Exposure) metric that shows the time spent on crash discovering. Experimental setup included one machine with two 32-core AMD EPYC 7542 CPUs, 512 GB of RAM, and Ubuntu 20.04 LTS. We compared our hybrid directed fuzzer (Sydr-Fuzz) with pure LibAFL-DiFuzz and other directed fuzzers, such as AFLGo, BEACON, WindRanger, WAFLGo, FishFuzz, and Prospector. For the evaluation, we selected five different projects from the AFLGo experimental set. These projects have publicly known CVEs in specific versions with TTE no greater than 150 minutes on AFLGo. The set of chosen target points with their IDs, locations, and experiment timeouts, is shown in Table 1.

*Table 1. Target points chosen for experiments.*

| Project | Target point ID | Location | TimeOut |
|---|---|---|---|
| cxxfilt | cxxfilt$_1$ | libiberty/cp-demangle.c: 1596 | 45 min |
| | cxxfilt$_2$ | libiberty/cplus-dem.c: 4319 | |
| giflib | giflib | util/gifsponge.c: 61 | 30 min |
| jasper | jasper | src/libjasper/mif/mif_cod.c:491 | 30 min |
| libming | libming | util/outputscript.c:1493 | 150 min |
| libxml | libxml$_1$ | valid.c:1181 | 60 min |
| | libxml$_2$ | valid.c:1189 | |

Each experiment was run several times with a defined timeout. In Sydr-Fuzz experiments, one instance each of LibAFL-DiFuzz and Sydr were launched simultaneously and worked in parallel. Sydr inverted branches in direct order with one solving thread. Every Sydr process was limited up to 2 minutes, with a 10-second timeout for solving a single SMT-query and a 60-second total solving time. Strong optimistic solutions [27] and symbolic address fuzzing [28] were enabled for Sydr. For each execution, we measured time intervals between the experiment start and different crashes discovering. After the end of experiment, we checked crash seeds with *gdb* to validate they were real crashes. Experiments results for the best attempts with outliers filtered are shown in Table 2. Columns represent different instruments, lines – different target points, and each cell contains TTE value in seconds. Empty cells mean that the instrument was not able to reach target point.

*Table 2.TTE experiments results.*

| TP ID | Sydr-Fuzz, s | LibAFL -DiFuzz, s | AFLGo, s | BEACON, s | WAFLGo, s | WindRanger, s | FishFuzz, s | Prospector, s |
|---|---|---|---|---|---|---|---|---|
| cxxfilt$_1$ | **22** | 175 | 49 | 60 | 41 | 44 | 384 | 470 |
| cxxfilt$_2$ | 136 | - | 63 | **55** | 3043 | 704 | 1208 | 347 |
| giflib | **3** | **3** | 18 | 4 | 5 | 55 | 41 | 45 |
| jasper | 18 | **16** | **16** | 50 | - | 81 | 287 | 22 |
| libming | 119 | 127 | 412 | **53** | - | - | 862 | 59 |
| libxml$_1$ | **8** | 14 | 28 | 43 | 107 | 666 | 2285 | 1694 |
| libxml$_2$ | 61 | 300 | 343 | **44** | - | 1108 | - | - |

TTE results show that Sydr-Fuzz hybrid directed fuzzer outperforms all other directed fuzzers on 3 out of 7 examples: **cxxfilt**$_1$, **giflib**, and **libxml**$_1$ with speedup up to 1.86 times. For **jasper** and **libxml**$_2$ Sydr-Fuzz results are second after the best measurements, and the difference is relatively small. As we see in Table 2, BEACON is another tool that has 3 out of 7 best results, however it supports only single-target mode, in opposite to all other tools. This can have direct influence on better TTE results, since in multi-target projects (**cxxfilt**, **libxml**) other tools spend their energy on reaching several target points simultaneously.

Comparing Sydr-Fuzz approach with pure LibAFL-DiFuzz, we can see improvements for all examples except **jasper**. This is caused by CPU load introduced by Sydr-Fuzz extra actions (Sydr launching, synchronization) since TTE for **jasper** is relatively short. Since the first synchronization with Sydr is performed after 60 seconds, improvements for **cxxfilt**$_1$, **giflib**, and **libxml**$_1$ can be mainly explained by the randomness of fuzzing process. However, for **cxxfilt**$_2$, **libming**, and **libxml**$_2$ hybrid approach shows a real advantage with speedup up to 4.9 times. What is more, Sydr-Fuzz has managed to reach target point **cxxfilt**$_2$ in a relatively short time, whereas pure LibAFL-DiFuzz has failed to reach it at all within 45 minutes timeout.

## *6. Conclusion*

In this paper, we propose a hybrid approach to directed fuzzing with novel seed scheduling algorithm, based on target-related interestingness and coverage. The approach also performs minimization and sorting of objective seeds according to the target-related information. We implement our approach in Sydr-Fuzz tool using LibAFL-DiFuzz as directed fuzzer and Sydr as dynamic symbolic executor.

We evaluate our approach with Time to Exposure metric and compare it with pure LibAFL-DiFuzz, AFLGo, BEACON, WAFLGo, WindRanger, FishFuzz, and Prospector. We get 3 out of 7 best results with speedup up to 1.86 times from the second-best results. We get significant improvements on 3 out of 7 examples comparing with pure LibAFL-DiFuzz fuzzer with speedup up to 4.9 times, and reaching an undiscovered target point. To sum up, Sydr-Fuzz hybrid approach to directed fuzzing shows high performance and helps to improve directed fuzzing efficiency.

## *Информация об авторах / Information about authors*

Дарья Алексеевна ПАРЫГИНА – магистр Московского государственного университета имени М.В. Ломоносова. Сфера научных интересов: символьное выполнение, гибридный фаззинг, направленный фаззинг.

Darya Alekseevna PARYGINA – master of Lomonosov Moscow State University. Research interests: symbolic execution, hybrid fuzzing, directed fuzzing.

Тимофей Павлович МЕЖУЕВ – бакалавр Московского государственного университета имени М.В. Ломоносова, лаборант Института системного программирования. Сфера научных интересов: символьное выполнение, гибридный фаззинг, направленный фаззинг.

Timofey Pavlovich MEZHUEV – bachelor of Lomonosov Moscow State University, laborant of the Institute for System Programming of the RAS. Research interests: symbolic execution, hybrid fuzzing, directed fuzzing.

Даниил Олегович КУЦ – кандитат технических наук, младший научный сотрудник Института системного программирования. Сфера научных интересов: динамический анализ, фаззинг, символьное выполнение, гибридный фаззинг.

Daniil Olegovich KUTZ – Cand. Sci. (Tech.), junior research assistant of the Institute for System Programming of the RAS. Research interests: dynamic analysis, fuzzing, symbolic execution, hybrid fuzzing.